\documentclass[conference]{IEEEtran}
\usepackage{algorithm}
\usepackage{algpseudocode}
\usepackage{placeins}
\usepackage{subcaption}
\usepackage{booktabs}
\usepackage{amsfonts}
\usepackage{graphicx}
\usepackage[dvipsnames,svgnames]{xcolor}
\usepackage{multirow} 

\ifCLASSINFOpdf
\else
\fi

\begin{document}
%\title{CONFLUX: Cache-Optimized Dynamic Sparse Attention for Long-Context LLM Prefill on FPGA}
\title{FAST-Prefill: \underline{F}PGA \underline{A}ccelerated \underline{S}parse \underline{At}tention for Long Context LLM Prefill}
%Dynamic-Sparse Attention Acceleration for Long-Context LLM Inference on FPGAs}

 \author{\IEEEauthorblockN{Rakshith Jayanth, Viktor Prasanna}
\IEEEauthorblockA{Ming Hsieh Department of Electrical and Computer Engineering\\
University of Southern Californa\\
Los Angeles, USA\\
Email: \{jayanthr, prasanna\}@usc.edu}
% \and
% \IEEEauthorblockN{Homer Simpson}
% \IEEEauthorblockA{Twentieth Century Fox\\
% Email: homer@thesimpsons.com}
% \and
% \IEEEauthorblockN{James Kirk\\ and Montgomery Scott}
% \IEEEauthorblockA{Starfleet Academy\\
% San Francisco, California 96678--2391\\
% Telephone: (800) 555--1212\\
%  Fax: (888) 555--1212}
}


%Telephone: (800) 555--1212, Fax: (888) 555--1212}


\maketitle

\begin{abstract}
In long-context large language model (LLM) inference, the prefill stage dominates computation due to self-attention over the complete input context. Sparse attention significantly reduces self-attention computation by limiting each token’s interactions to a subset of tokens. The attention sparsity pattern varies across input prompts, and within a prompt, each attention head can follow a distinct pattern. This makes attention sparsity dynamic. The requirement of generating the sparsity pattern, combined with limited data reuse in attention, shifts the prefill compute to being memory-bound. 
%Long-context LLM inference with dynamic sparse attention shifts the prefill-stage compute to become memory-bound as opposed to dense long-context inference, which is compute-bound. 
%This arises from two factors: (1) generation of sparse Key-Value (KV) cache vector indices for sparse attention, which majorly involves element-wise operations and other bandwidth-heavy operations such as sorting, and (2) sparse attention, which involves frequent irregular access ytof KV cache vectors. 
This, in addition to the huge energy requirements for long-context inference on GPU, motivates FPGAs as good candidates for accelerating dynamic long-context inference. However, acceleration on FPGA faces the following key challenges: \textit{(1) Generation of large intermediary tensors during sparse index generation, (2) Sparsity pattern dependent KV cache access limits the effectiveness of naive prefetching in addition to bandwidth under-utilization, and (3) Limited DSP resources for high-throughput matrix multiplication.}   

To tackle these challenges, we propose FAST-Prefill, the first FPGA accelerator for long-context prefill-stage inference with dynamic sparse attention. To efficiently generate sparse indices, we propose a \textit{fused pipeline unit with a memory-aware execution order} to reduce large tensors and irregular memory accesses. To reduce off-chip memory traffic for accessing the KV cache, we utilize the memory hierarchy to design a \textit{liveness-driven, dual-tier cache}. For high-throughput matrix multiplication, we design a \textit{hybrid Matrix Processing Unit (MPU)} with DSPs and bit-plane decomposition using LUTs. 
%In order to limit the usage of DSP resources and increase matrix-multiplication throughput, we design multi-lane bit-plane based matrix-multiplication unit. 
We implement FAST-Prefill on Alveo U280 and evaluate it on the Llama and Qwen models (batch size = 1) for context lengths ranging from 4K to 128K tokens. We demonstrate an average speedup of up to 2.5$\times$ in TTFT and 4.5$\times$  improvement in energy efficiency over GPU implementation on Nvidia A5000 GPU. 
%Large language models (LLMs) are increasingly deployed in settings demanding long contexts. Long-context LLM inference is bottlenecked by the quadratic computational complexity of self-attention in the prefill stage. Sparse self-attention, where each token interacts with only a subset of tokens, is a common approach to reduce this complexity. Still, executing long-context inference on resource-constrained platform such as FPGA, however, is challenging due to the following. (1) High cost in generating sparse indices due to softmax-based ranking, (2) Large memory traffic due to frequent irregular KV cache access, and (3) High DSP pressure due to a large number of parallel dot-products.

%This paper proposes FAST-Prefill, a novel FPGA accelerator for efficient long-context LLM inference. FAST-Prefill features (1) Max-based logit space ranking using LUTs for selecting sparse indices, (2) custom hardware cache for selective caching blocks, KV cache to significantly reduce memory traffic, and (3) Bit-plane LUT-based dot-product, which significantly relieves DSP pressure along with facilitating large parallel matmuls. 
%Absence of control over memory hierarchy in GPUs and inefficient computation of not-so parallelisable heavy kernels in sparse attention, makes FPGAs a potential candidate for efficient sparse long-context LLM inference.     
\end{abstract}


% \fi
\IEEEpeerreviewmaketitle
\section{Introduction}
Large language models (LLMs) have gained significant popularity in recent years for their exceptional performance across a wide range of natural language processing (NLP) tasks. Specifically, long-context inference is critical in tasks such as document summarization \cite{liao2025e2llm, bai2025longbench}, code generation \cite{huynh2025large, yang2025empirical}, and Q\&A \cite{yan2025mir,xiao2025generalizing}. However, long-context inference incurs a huge computation cost due to self-attention that grows quadratically with context length (prompt length) in the prefill phase of LLM inference \cite{vaswani2017attention}. Recent works \cite{NEURIPS2020_c8512d14, NEURIPS2024_5dfbe6f5, xiao2024efficient,hao2025omnikv,flexprefill} target sparse attention, in which each token interacts with only a subset of tokens, resulting in a drastic reduction in computational requirements. Unlike weight compression and sparsification techniques such as N:M sparsity \cite{NEURIPS2021_6e8404c3} and block sparsity \cite{9424344}, which sparsify weight matrices, attention sparsity results in sparse attention scores during self-attention. State-of-the-art works \cite{flexprefill,NEURIPS2024_5dfbe6f5} target dynamic sparsity in self-attention wherein the sparsity pattern varies for each input prompt. Additionally, for a given prompt, each attention head can exhibit a unique pattern.

%These attention sparsity patterns are dynamic; that is, they vary across the attention heads and input prompts. 
Current works target GPUs for evaluating sparse attention algorithms due to the high computational capabilities they offer. However, sparse attention makes the inference memory-bound. This arises from two key factors. (1) Generating a specific set of tokens for each token (Sparse index generation) requires a data-dependent control flow with low compute intensity. This results in under-utilized compute resources and saturates the memory bandwidth on the GPU. 
(2) Sparse attention requires each query to access a different subset of Key-Value (KV) vectors. Unlike dense attention where all queries access the complete KV cache \cite{pope2022efficientlyscalingtransformerinference}, sparse patterns cause each query to read different portions of the KV cache, preventing efficient data reuse.
%(2) Sparse attention involves sparse pattern-dependent access of KV cache vectors as each query (Q) vector interacts with a different set of KV vectors. This deviates from the GPU's preference of sequential access, resulting in inefficient menory transactions. FPGAs facilitate the design of custom kernels for element-wise operations and make efficient use of the memory hierarchy (HBM-URAM-BRAM), which motivates accelerating dynamic sparse attention-based long-context inference on FPGAs.
Additionally, the high energy requirements of GPUs drive the consideration of energy-efficient platforms such as FPGAs. However, there has been limited exploration towards accelerating long-context LLM inference with dynamic attention sparsity on FPGAs.

Accelerating long-context LLM inference on an FPGA, however, faces substantial design challenges. \textbf{(1)} The naive design for sparse index generation results in large intermediary tensors. On-chip buffers are insufficient to hold these, and off-chip storage results in frequent access, contributing to increased latency. \textbf{(2)} The large size of the KV cache ($\sim$3-4 GB) restricts it to off-chip storage. Ensuring efficient reuse of KV vectors across different queries is non-trivial, as the access is dependent on the sparsity pattern as opposed to naive streaming. Additionally, fetching vectors too early or too late can result in wasted bandwidth or compute stalls. \textbf{(3)} Long context inference involves multiple matrix multiplication operations. Current FPGA-based works \cite{flightllm, flightvgm} primarily rely on DSP-based multiplication. This bottlenecks the design as multiple matrix operations become largely sequential. We detail the challenges in Section-\ref{Sec:challenges}.

\begin{figure}[t]
    \centering
    \includegraphics[width=\linewidth]{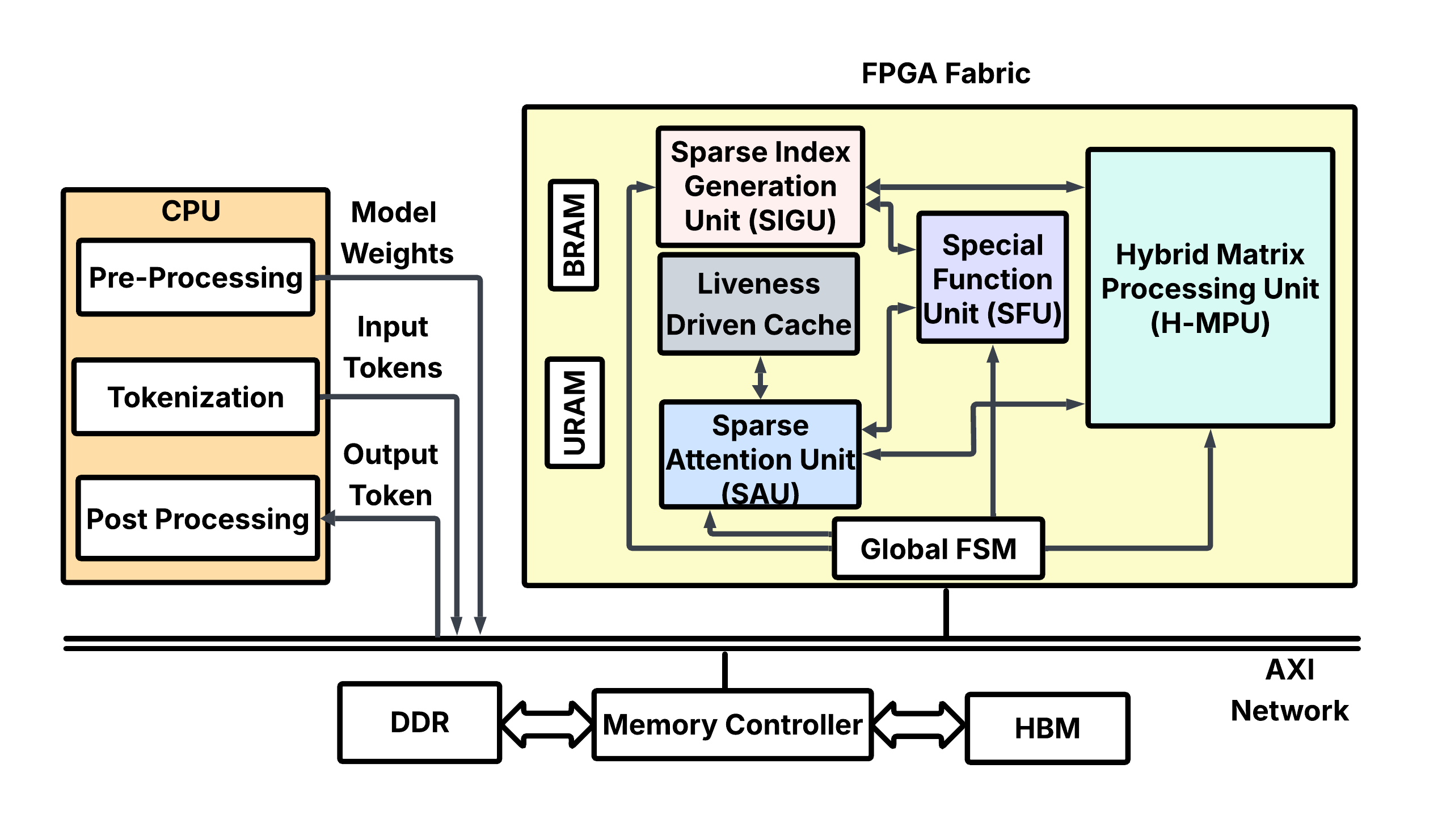}
    \caption{FAST-Prefill Architecture}
    %\vspace{-0.2cm}
     \label{fig:system_workflow}
\end{figure}

To address these challenges, we propose FAST-Prefill, an FPGA-based accelerator for efficient long-context LLM inference with dynamic sparse attention. The main contributions of the paper are summarized as follows:
\begin{itemize}
    \item To the best of our knowledge, we propose the first FPGA accelerator that targets dynamic sparse prefill-stage inference for long-context LLMs with W8A8 precision.
    \item We propose a \textbf{streaming, memory-aware sparse index generation architecture} that computes block-level relevance scores through incremental aggregation, eliminating the need to generate large intermediate tensors. This transforms index generation from a generate-then-aggregate operation requiring $\sim$4GB of intermediate storage, into a stream-and-accumulate operation requiring with $\sim$4KB storage. 
    %\item We propose a custom sparse-index generation kernel to address Challenge-1. CONFLUX performs memory-system-aware restructuring of computation order to stream Key blocks in mostly sequential bursts and fuses the index generation pipeline to limit intermediate tensors.
    \item We propose a \textbf{liveness-driven custom cache} to store a small set of KV cache blocks. This analyzes sparse index patterns to prefetch KV cache blocks from off-chip memory (HBM) via efficient coordinated bursts.
    \item The custom cache is \textbf{dual-tier} with a threshold-based admission policy that places high-reuse blocks of KV cache in the Hot tier and low-reuse blocks in the Cold tier to prevent cache thrashing.
    %\item We propose an access-frequency-based custom cache with block-granular KV cache residency management to address Challenge-2. The cache consists of two separations for Hot and Cold KV blocks, with each segregation having a dedicated replacement policy based on predicted future use.
    \item We propose a hybrid Matrix Processing Unit (MPU) that includes systolic array grids for bit-plane matrix multiplication using LUTs, in addition to DSP-based systolic array grids. 
    %\item We perform multi-lane bit-plane matrix multiplication to address Challenge-3. This facilitates parallel matrix multiplication with no DSP pressure. CONFLUX features a scheduler to schedule the attention block computation to each matmul array, based on their accessibility, to limit idling of the arrays.
    \item Our implementation of FAST-Prefill on Xilinx Alveo U280 achieves up to 1.2-2.5$\times$ improvement in Time To First Token (TTFT) over the implementation of dynamic sparse attention on Nvidia RTX A5000 GPU across context lengths of 4K-128K tokens. We also achieve 4.5$\times$ energy efficiency over the GPU.
\end{itemize}

\begin{figure*}[t]
    \centering
    \includegraphics[width=0.99\textwidth]{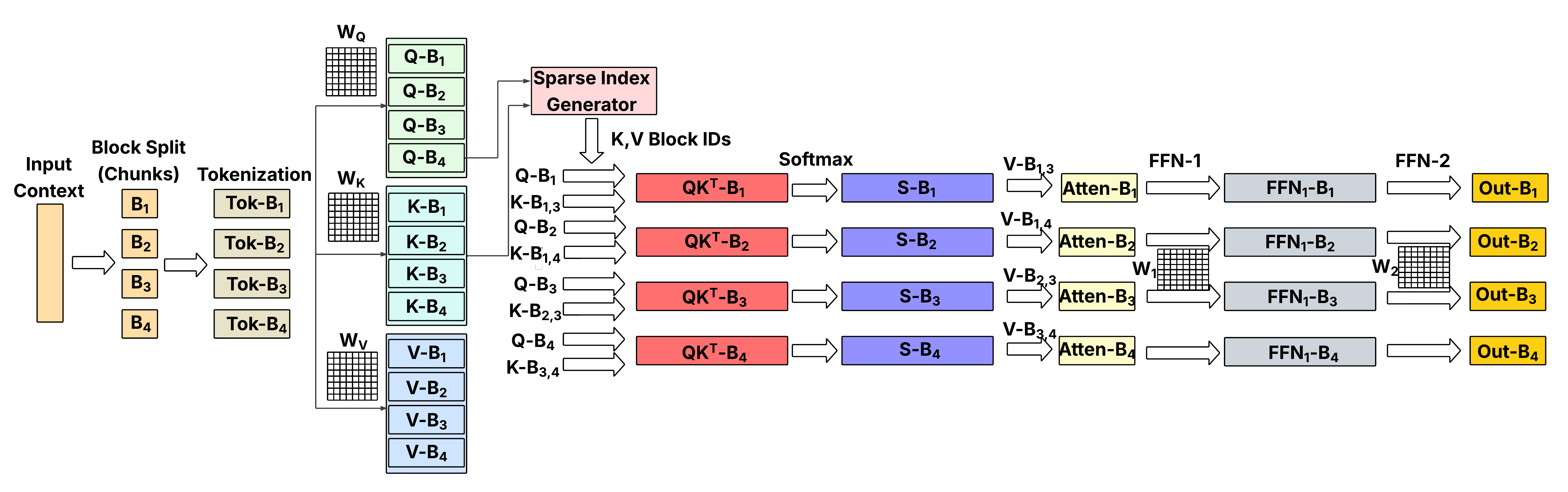}
    \caption{Prefill Workflow with Sparse Attention}
    \vspace{-0.2cm}
     \label{fig:sparse_attention_prefill}
\end{figure*}

\section{Background}
\subsection{LLM Inference}
\label{Sec_LLM_Inference}
LLM inference consists of two compute stages: the prefill stage and the decode stage. In the prefill stage, the complete input sequence (X) is tokenized (E) and the Key-Value (KV) cache along with Query (Q) matrix is generated as shown in Equation-\ref{eq_QKV_dense}.   
\begin{equation}
Q = W_{Q}.E(X),\quad K = W_{K}.E(X), \quad V = W_{V}.E(X)
\label{eq_QKV_dense}
\end{equation}
where $W_Q$, $W_K$, and $W_V$ are Multi-Head self-attention (MHA) weight matrices.  
Using QKV matrices, causal MHA is computed according to Equation-\ref{eq_Attn_dense}. 
\begin{equation}
 A = Softmax(\frac{Q.K^T}{\sqrt{d}})V
\label{eq_Attn_dense}
\end{equation}
The MHA output is further processed by the Feed Forward Network (FFN) which consists of two fully-connected linear layers along with a non-linear activation function. The FFN is computed as shown in Equation-\ref{eq_FFN_dense}.
\begin{equation}
 O = g(AW_1)W_2
\label{eq_FFN_dense}
\end{equation}
where, $W_1$ and $W_2$ are FFN weights and $g$ is the non-linear activation function.
The output $O$ of FFN is context embedding for the next layer.
The prefill stage concludes with the generation of the first output token. During the decode stage, the output tokens are generated auto-regressively i.e., one token in each model iteration. The compute in decode stage follows the same equations as prefill but only with the new token generated at the end of previous iteration. This transforms the matrix-matrix multiplications of prefill stage into matrix-vector multiplications during decode stage.   

\subsection{Long-Context LLM Inference: Sparse Attention}
\label{Sec_Long_context_LLM}
During prefill stage, the computation complexity of self-attention grows quadratically  ($O(S^2)$) with context length ($S$). For long contexts, this explodes the latency for generating the first token (TTFT) due to large compute and data transfers. To mitigate this, prior works on long-context inference consider sparse attention during the prefill stage, where each token is required to interact with only a subset of tokens. State-of-the-art works target dynamic sparse attention, wherein for each attention head in MHA, tokens can follow a unique sparsity pattern. We leverage the state-of-the-art sparse index generation algorithm of Flex-Prefill \cite{flexprefill}. We detail the algorithm for sparse index generation in Algorithm-\ref{alg:sparse_idx_gen}.

\begin{algorithm}
\caption{Sparse Index Generation (Each Attention Head): Flex-Prefill}
\textbf{Input:} Q, K, V, $\tau$, $\gamma$  \\
\textbf{Output:} Sparse index set, S 
\label{alg:sparse_idx_gen}
\begin{algorithmic}[1]
\State{select $\hat{Q}$ = $Q_{[-block\_size:]}$}
{\color{blue}\Comment{Representative query subset (last block)}}
\State{$\bar{a}$ $\leftarrow$ softmax(pool($\hat{Q}$)pool($K$)$^T$/$\sqrt{d}$)}
{\color{blue}\Comment{Estimated block-pooled attention}}
\State{$\hat{a}$ $\leftarrow$ pool(softmax($\hat{Q}K^T$/$\sqrt{d}$))}
{\color{blue}\Comment{True block-pooled attention}}
\State{$d_{JS}$ $\leftarrow$ $\sqrt{JSD(\bar{a}||\hat{a})}$} 
{\color{blue}\Comment{Jensen-Shannon divergence}}
\If{ $d_{JS} < \tau$} {\color{blue}\Comment{Determine Pattern}}
\State{pattern $\leftarrow$ query\_specific}
\Else
\State{pattern $\leftarrow$ vertical\_slash}
\EndIf
%\State{\Return pattern}
\Statex{\textbf{Pattern == Vertical\_Slash}}
\State{$\hat{a}$ $\leftarrow$ pool(softmax($\hat{Q}K^T$/$\sqrt{d}$)), $where \hat{Q} \subset Q$} 
{\color{blue}\Comment{Compute a subset of the full attention map}}
\State{$a_v$ $\leftarrow$ sum\_vertical($\hat{A}$)/$\Sigma_{i,j}\hat{A}[i,j]$} 
{\color{blue}\Comment{ Sum \& normalize attention scores along the vertical directions}}
\State{$a_s$ $\leftarrow$ sum\_slash($\hat{A}$)/$\Sigma_{i,j}\hat{A}[i,j]$}
{\color{blue}\Comment{ Sum \& normalize attention scores along the slash directions}}
\State{$I_v \leftarrow$ argsort($a_v$)} {\color{blue}\Comment{Sort vertical and slash attention scores}}
\State{$I_s \leftarrow$ argsort($a_s$)} 
\State{$K_v \leftarrow$ min(\{$k : \Sigma_{i\in I_v[1:k]}a_v[i] \geq \gamma$\}} {\color{blue}\Comment{minimum computational budget making the sum of the scores exceeds $\gamma$ }}
\State{$K_s \leftarrow$ min(\{$k : \Sigma_{i\in I_s[1:k]}a_s[i] \geq \gamma$\}} 
\State{$S_v \leftarrow I_v[1:K_v]$} {\color{blue}\Comment{Select indices}}
\State{$S_s \leftarrow I_s[1:K_s]$}
\State{$S \leftarrow S_v \cup S_s$}
\State{\Return S}
\Statex{\textbf{Pattern == Query\_Aware}}
\State{$\bar{Q} \leftarrow $ pool(Q), $\bar{K} \leftarrow $ pool(K)} {\color{blue}\Comment{Compute estimated attention scores using pooled Q and K}}
\State{$\bar{A} \leftarrow $ softmax($\bar{Q}\bar{K}^T$/$\sqrt{d}$)}
\State{$\bar{A} \leftarrow $ flatten($\bar{A}$/$\Sigma_{i,j}\bar{A}[i,j]$)} {\color{blue}\Comment{Flatten and normalize attention map}}
\State{$I_a \leftarrow$ argsort($\bar{A}$)} {\color{blue}\Comment{Sort attention scores}}
\State{$K \leftarrow$ min(\{$k : \Sigma_{i\in I_a[1:k]}\bar{A}[i] \geq \gamma$\}} {\color{blue}\Comment{minimum computational budget making the sum of the scores exceeds $\gamma$ }}
\State{$S \leftarrow I_a[1:K]$} {\color{blue}\Comment{Select indices}}
\State{\Return S}
% \State{Lat = []}
% \For{each Proposer \textit{p}}
% \State{Lat[p] = get-maximum-raw-latency()}
% \EndFor
% \State{Sort(Lat, Proposers, order = Decreasing)} 
% \Comment{Initial order}
% \For{each compute unit \textit{cu\_i}}
% \State{pf\_free[d] = 0.0}
% \Comment{Time after which the Prefetch thread is free}
% \State{dev\_free[d] = 0.0}
% \Comment{Time after which the Device (compute) thread is free}
% \EndFor
% \For{each Proposer \textit{p} in sorted order}
% \For{each compute unit \textit{cu\_i}}
% \State{gen\_s = get-est-generation-lat(p,i)}
% \State{prefetch\_s = get-est-fetch-lat(p,i)}
% \State{import\_s = get-est-import-lat(p,i)}
% \State{t\_pf = pf\_free[i] + prefetch\_s}
% \State{t\_imp0 = max(dev\_free[d], t\_pf)} 
% \State{t\_imp1 = t\_imp0 + import\_s}
% \State{t\_gen[i] = t\_imp1 + gen\_s}
% \EndFor
% \State{Map \textit{p} to compute unit \textit{i*} with minimum t\_gen[i*]}
% \State{pf\_free[i*] = t\_pf(i*)}
% \State{dev\_free[i*] = t\_gen(i*)}
% \EndFor
\end{algorithmic}
\end{algorithm}

During self-attention, each attention head generates an attention matrix of size $S\times S$. Sparse attention views the attention matrix as a block matrix with each block of size $B\times B$ where $B$ is generally 64 or 128, computed from $B$ query vectors and $B$ key vectors. As part of sparse attention, a few blocks are selected corresponding to a pattern in each attention head, and attention computation takes place for these specific Q, K, and V blocks. 
Flex-Prefill algorithm classifies each attention head to follow a query-specific sparsity pattern or a fallback pattern of vertical-slash \cite{NEURIPS2024_5dfbe6f5}. For an attention head that follows vertical-slash pattern, the blocks selected will be along the diagonals or columns of the attention matrix. For attention-heads that follow query-specific pattern, the selected blocks are distributed across different rows and columns of attention matrix. Flex-Prefill identifies each attention head to follow either of the two types of sparsity patterns by performing dense attention on the last query blocks and computing the Jensen-Shannon Divergence. This value is compared against a threshold $\tau$ (set to 0.1). This threshold estimates the extent of adverse impact on output quality when query-aware pattern is opted for the given attention head. If the divergence is less than the threshold, the query-aware pattern is selected; else, the conservative vertical-slash pattern is selected.
For each attention head, the algorithm returns indices of the corresponding Key matrix blocks (B = 128 tokens) for each Query block (128 tokens). Based on the sparse index set $S$ for each attention head, the attention is computed as follows:
\begin{equation}
     A = Softmax(\frac{Sparse\_Attention(Q.K^T, S)}{\sqrt{d}})V
\label{eq_Attn_sparse}
\end{equation}

%\subsection{DYnamic Sparse Attention}

\section{Challenges}
\label{Sec:challenges}

%{Accelerating long-context LLM inference on an FPGA, however, faces substantial design challenges.} \\
In this section, we detail the challenges in accelerating long-context LLM inference on FPGA.\\
\textbf{\textit{Challenge-1: Sparse index generation generates large intermediary tensors combined with irregular memory access to Key (K) vectors.}}

Computing sparse indices requires scoring all Key blocks against the last query-block vectors to identify the top-K most relevant blocks. Considering a context length of 128K tokens, this can produce 128$\times$128K per attention head, resulting in more than 2GB of intermediate storage. Additionally, operations such as softmax, exponentials, and block pooling produce 2GB of intermediate tensors. This is prohibitively large for on-chip storage, and frequent off-chip access explodes latency. Standard fusion techniques are not effective because they primarily fuse element-wise operations rather than operations with different dimensions.\\    
% and LLama3.2-3B configuration, this will produce  
% Sparse index generation involves interaction with specific windows of Key vectors across different attention heads. This is essential to generate sparse indices specific to each attention head. However, this results in accessing different sets of Key vectors (K-blocks), leading to repeated non-contiguous reads. This contributes to multiple short bursts and hence, poor DDR/HBM bandwidth efficiency. Additionally, during index generation, large intermediary tensors are generated which if written to and read from the off-chip memory, will exacerbate memory traffic, exploding latency. \\
\textbf{\textit{Challenge-2: Sparsity pattern dependent KV cache access.}} 
% Sparse attention with GQA (32 query heads, 8 KV head caches) creates 
% irregular access patterns: each query head selects 64 different blocks 
% from ~1024 available blocks, with patterns varying across heads. This 
% creates several problems:

% (1) UNPREDICTABLE PREFETCH: It's unclear which blocks to prefetch and when
%     - Each head needs different blocks
%     - Access order depends on runtime sparse indices
%     - Naive prefetching wastes limited URAM capacity

% (2) BANDWIDTH FRAGMENTATION: Fetching blocks on-demand creates:
%     - Many small 16 KB HBM reads scattered across the address space
%     - Underutilized HBM bandwidth (short bursts)
%     - Pipeline stalls waiting for memory

% (3) MISSED REUSE OPPORTUNITIES: Blocks may be used by multiple heads
%     within the same KV head group, but without coordination, each head
%     fetches independently, wasting bandwidth.
Reuse of KV cache blocks is dependent on the sparsity pattern corresponding to each attention head. This, neither purely streaming nor purely temporal, access pattern of KV cache blocks results in the following issues. (a) Naive prefetching along the memory hierarchy is ineffective as prefetch timing is critical. Prefetching a KV block too early can waste the limited on-chip resources, and prefetching the block too late results in pipeline stalls. (b) Fetching blocks on demand leads to many small off-chip memory reads, resulting in underutilized bandwidth and pipeline stalls. (c) Limiting the reuse of KV blocks across attention heads in the case of Group-Query-Attention (GQA) leads to independent fetching of the blocks, which increases memory traffic.\\
% Reuse of KV cache vectors (blocks of KV cache) is dependent on the sparsity pattern corresponding to each attention head. This neither purely streaming nor purely temporal access pattern of KV cache blocks makes simple prefetching along the memory hierarchy (HBM-URAM-BRAM) ineffective and can also adversely impact bandwidth. Additionally, prefetch timing is also critical. Prefetching a KV block too early can waste the limited on-chip resources, and prefetching the block too late results in pipeline stalls. \\
\textbf{\textit{Challenge-3: Constrained Matrix multiplication throughput due to limited DSPs.}} 
Long-context inference results in multiple block-sized matrix multiplications across various stages in the pipeline (QKV, Sparse Attention and FFN). Using DSPs limits the matrix multiplication architecture to about six 32x32 systolic arrays on U280 FPGA, insufficient for the large number of matmuls involved.

\section{Design Methodology}
%\subsection{CONFLUX Overview}

\subsection{Hardware Architecture Overview} 
% \sloppy
We design an efficient, high-performance streaming data flow accelerator for dynamic-sparsity-based long-context LLM inference. Figure~\ref{fig:system_workflow} shows the overall architecture of FAST-Prefill. The major components in FAST-Prefill include a Global Finite State Machine (FSM) (Section~\ref{Sec:other_units}), Special Function Unit (SFU)(Section~\ref{Sec:other_units}), and multiple compute units. The primary compute units include the Sparse Index Generation Unit (SIGU) (Section~\ref{Sec_SIGU}), the Sparse Attention Unit (SAU) (Section~\ref{Sec_SAU}), and a Hybrid Matrix Processing Unit (MPU) (Section~\ref{Sec_MPU}).

% We design an efficient, high-performance streaming data flow accelerator for 
% % dynamic sparsity-based long-context LLM inference. 
% % Fig~\ref{fig:system_workflow}  FAST-Prefill. The major components include a 
% Global Finite State Machine (FSM) (Section~\ref{Sec:other_units}), a 
% Special Function Unit (SFU) (Section~\ref{Sec:other_units}), and multiple compute units. The primary compute units are the Sparse Index Generation Unit (SIGU) 
% (Section~\ref{Sec_SIGU}), the Sparse Attention Unit (SAU) (Section~\ref{Sec_SAU}), and the Hybrid Matrix Processing Unit (MPU) (Section~\ref{Sec_MPU}).

Given the large number of input tokens to be processed, we adopt chunked prefill \cite{agrawal2023sarathiefficientllminference}, wherein the input context is split into chunks of equal token length. We set the chunk size to 128 tokens to align with the block granularity used in FlexPrefill. In accordance with algorithmic semantics, chunks are streamed through the pipeline; however, for stages that require global context—such as sparse index generation—explicit barrier buffers are inserted to ensure correctness. The CPU performs tokenization and context embedding on the input prompt and transfers the token embeddings to HBM. Subsequent computation proceeds entirely on the FPGA in a streaming fashion across LLM layers till the generation of the first token. This marks the completion of the prefill stage.

For each transformer layer, chunks are first processed to generate Key and Value tensors, which are written to off-chip memory. Once all chunks for the layer are available, the SIGU is activated to compute sparse block indices based on the most recent query block and the complete set of Key blocks. These indices define the dynamic sparsity pattern for attention and are forwarded to the SAU. The SAU then executes sparse attention by selectively fetching the required KV blocks from memory, performing dot-product attention using the hybrid MPU, and accumulating the results. The attention output is streamed to compute matrix multiplication and activation functions as part of the feed-forward network (FFN).

%Given the large number of input tokens to be processed, we adopt the technique of chunked prefill wherein, the input context is split into chunks of equal token length. We set the chunk size to be 128 to align with the block size of 128 tokens in Flex Prefill. In accordance with the algorithm semantics, we pipeline the chunks wherever feasible or appropriate barrier buffers are inserted in the pipeline to wait until all chunks have been processed. Figure-b shows the end-to-end pipeline. The CPU performs tokenization and context embedding over the input prompt and transfers them to HBM. 
%Explain complete data flow.
%FSM??
\begin{figure}[t]
    \centering
    \includegraphics[width=\linewidth]{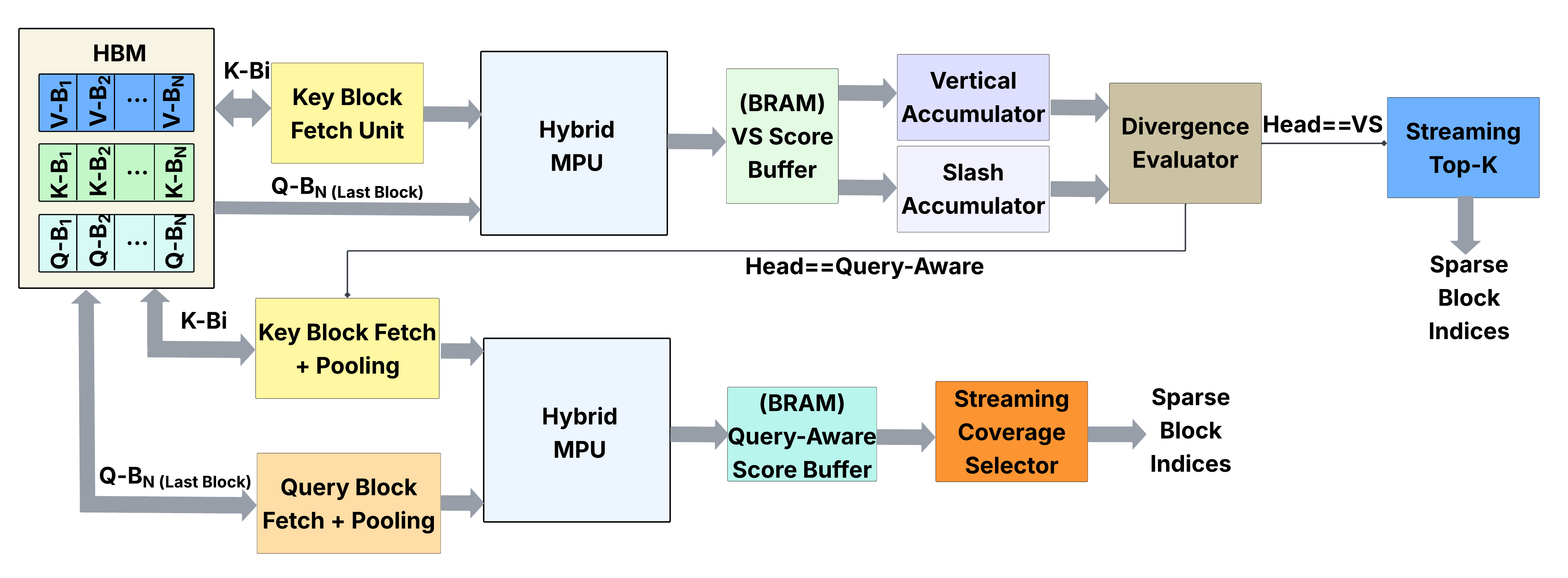}
    \caption{Sparse Index Generation Unit Workflow}
    %\vspace{-0.2cm}
     \label{fig:sparse_index_generation}
\end{figure}

\subsection{Attention Sparse Index Generation Unit}
\label{Sec_SIGU}
Sparse index generation in FlexPrefill determines, for each attention head, a subset of Key–Value (KV) blocks that are most relevant to the most recent query window. Concretely, given a sequence length $S$, head dimension $d$, and block size $B$ (128 tokens in our design), the algorithm operates on the last $B$ query vectors $\hat{Q} \in \mathbb{R}^{B \times d}$ and the full Key matrix $K \in \mathbb{R}^{S \times d}$. A naïve implementation explicitly computes intermediate attention tensors such as $\hat{Q}K^T \in \mathbb{R}^{B \times S}$, pooled attention maps of size $\lceil S/B \rceil$ $\times$ $\lceil S/B \rceil $, and per-head score vectors spanning all blocks. For long contexts (e.g., $S$=128K), these tensors are large—on the order of tens of megabytes per head —and their construction requires repeated, non-contiguous accesses to different Key blocks across heads. This results in multiple short HBM bursts, poor bandwidth utilization, and excessive off-chip traffic when intermediate tensors are written out and read back, significantly increasing latency. 

FAST-Prefill addresses this challenge by implementing sparse index generation as a streaming, memory-system-aware pipeline that avoids materializing large intermediate tensors and converts irregular Key access into mostly sequential bursts. The unit features a Key Block Fetch Unit, which retrieves Key blocks of size $B \times d$ from HBM and then into a small on-chip Key Block Buffer implemented in URAM. Crucially, Key blocks are fetched strictly in increasing block order, independent of attention head, ensuring that off-chip accesses occur as long, contiguous bursts rather than head-dependent random reads. 
As per Algorithm-\ref{alg:sparse_idx_gen}, the most recent query window $\hat{Q}$ (last 128 Q entries) is loaded once into a reused. 
For each streamed Key block, a Hybrid MPU performs dot products between $\hat{Q}$ and the current block using fixed-point arithmetic. Instead of storing the full $\hat{Q}K^T$ tensor, the design incrementally updates compact per-block statistics, thereby collapsing a $B \times S$ tensor into vectors of length $\lceil S/B \rceil$.

These per-block statistics are accumulated in on-chip Score Buffers (BRAM), which store block-level scores for each head. Two types of scores are maintained: vertical scores, which aggregate attention values column-wise (i.e., across queries for each Key block), and slash scores, which aggregate attention values along diagonals. The accumulation is performed by dedicated Vertical and Slash Accumulator units that update the score buffers as each Key block streams through, eliminating the need to revisit Key data. To determine which sparsity pattern to apply, the design computes a per-head divergence metric between two pooled attention distributions: one derived from block-pooled $\hat{Q}K^T$ and another from pooled $\hat{Q}$ and pooled $K$. This divergence, implemented in the Divergence Evaluation module, is compared against a threshold $\tau$ to select between query-aware and vertical-slash patterns. Operations in this module are element-wise or reduction-based and are implemented using LUT-based arithmetic and comparators.

Once block-level scores are available, the index selection stage enforces the coverage constraint, defined as selecting the smallest number of blocks whose cumulative attention value exceeds a user-defined threshold $\gamma$. Rather than performing a full argsort over all blocks—which would require random memory access and complex control—the design uses a Streaming Top-k Selection Module that processes score buffers sequentially. This module maintains a small, fixed-size candidate list and incrementally inserts blocks based on score comparisons until the cumulative sum crosses $\gamma$. The selection logic is implemented with comparator trees, prefix adders, and small LUT-based priority structures, yielding bounded latency and avoiding global sorting. 
%Because scores are already stored per block, selection operates entirely on BRAM-resident data. The final output is a compact list of block indices (not individual token indices), which is written once to off-chip memory and reused by the subsequent sparse attention stage.

%When the divergence metric falls below the threshold $\tau$, the kernel switches to the query-aware sparsity pattern, which directly selects Key blocks based on their relevance to the pooled query representation rather than relying on structural vertical or diagonal aggregation.
For attention heads following query-aware sparsity, the Query Pooling Module computes a block-level summary $\bar{Q} \in \mathbb{R}^{\lceil S/B \rceil \times d}$ by averaging query vectors within each block, while the Key Pooling Module similarly computes $\bar{K} \in \mathbb{R}^{\lceil S/B \rceil \times d}$ as Key blocks stream through the pipeline. Unlike the vertical-slash path, which requires maintaining separate vertical and diagonal accumulators, the query-aware path computes a single block-level attention score vector $\bar{A} = softmax(\bar{Q}\bar{K}^T/\sqrt{d})$, whose length is equal to the number of blocks. This computation is performed incrementally as each Key block is fetched, without materializing the full $\bar{Q}\bar{K}^T$ matrix.

The Query-Aware Scoring uses the Hybrid MPU for computing matrix multiplication on the pooled vectors. Scores are accumulated into a dedicated Query-Aware Score Buffer in BRAM, and normalization is applied using LUT-based exponential approximation followed by a running sum and reciprocal, avoiding floating-point softmax units. Selection of active blocks is governed by the coverage constraint $\gamma$: the Streaming Coverage Selector scans the query-aware score buffer in descending score order, accumulating attention values until the threshold is met. As in the vertical-slash case, this avoids a full argsort by using bounded, comparator-based selection logic and terminates early once sufficient coverage is achieved. The resulting set of block indices is directly emitted as the sparse index set $S$ for the head.

%Control over this process is provided by a dedicated Index Generation FSM, which orchestrates Key block streaming, per-head score accumulation, divergence evaluation, and index selection. The FSM ensures that all heads observe the same Key streaming order, thereby maximizing data reuse, while head-specific decisions (query-aware versus vertical-slash) are resolved locally using on-chip state. 
SIGU writes no intermediate tensor larger than $O(\lceil S/B \rceil)$ elements to off-chip memory, and each Key block is fetched exactly once. By fusing score computation, pooling, divergence evaluation, and top-k selection into a single streaming pipeline, FAST-Prefill transforms sparse index generation from an irregular, memory-bound operation into a deterministic, bandwidth-efficient datapath that preserves Flex-Prefill semantics while fully exploiting FPGA memory hierarchy.

\begin{figure}[t]
    \centering
    \includegraphics[width=0.99\linewidth]{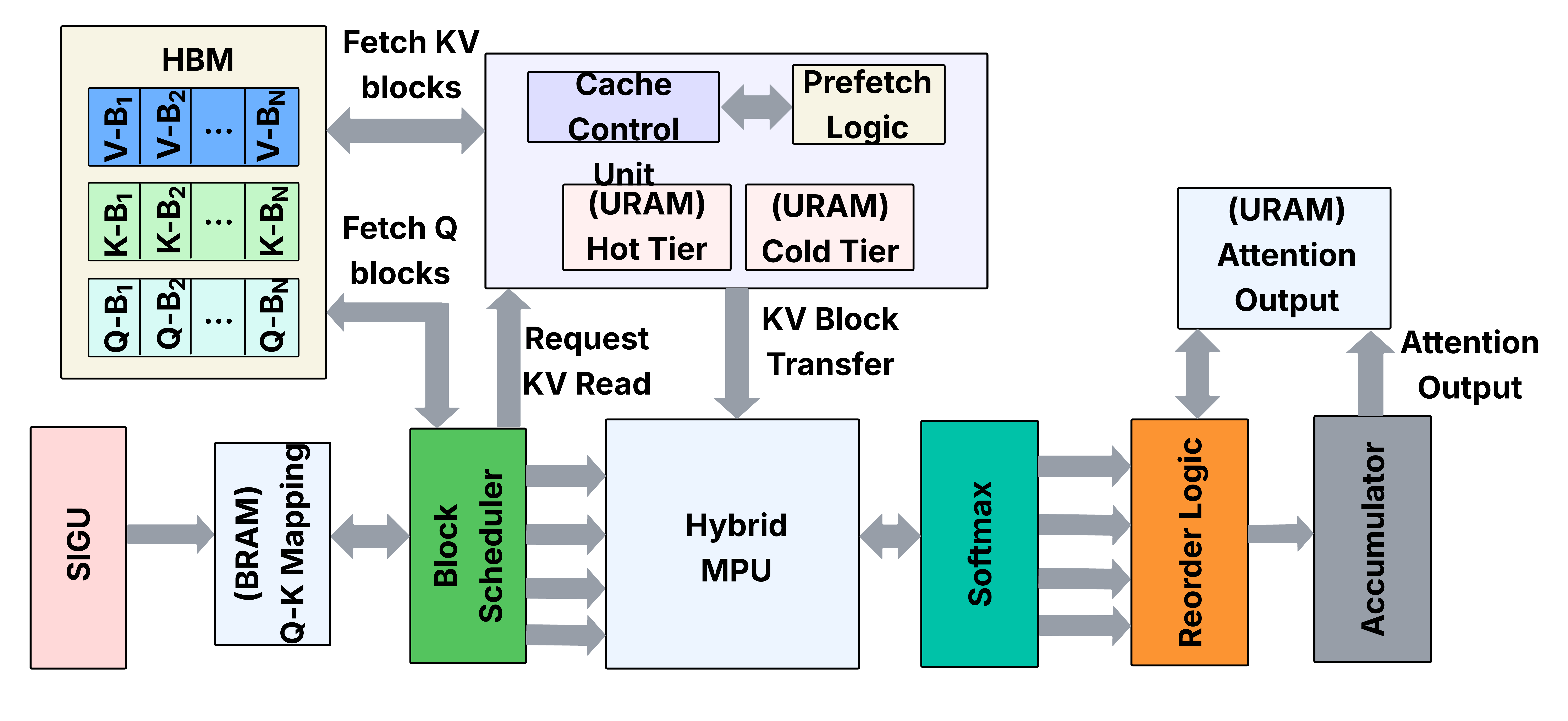}
    \caption{Sparse Attention Unit Workflow}
    \vspace{-0.2cm}
     \label{fig:sparse_attention_unit}
\end{figure}

\subsection{Sparse Attention Unit}
\label{Sec_SAU}

The sparse attention unit in FAST-Prefill integrates dynamic sparse attention computation with a custom cache (in URAM) designed to hold selective KV cache blocks. In addition to the Q, K and V vectors, the SAU also uses the sparse indices generated by the SIGU. 
The unit operates on Key and Value tensors stored in off-chip HBM as contiguous blocks of fixed size (128 tokens per block), and consumes a sparse index set that specifies, for each attention head and query block, which KV blocks participate in attention. Although the KV blocks themselves are stored contiguously, the sparse index set induces a head-dependent gather pattern: different heads and query blocks request different subsets of KV blocks, with potential overlap in blocks corresponding to the same attention group as per Group-Query-Attention. Executing these requests directly would result in repeated non-contiguous memory reads, each generating short HBM bursts and underutilizing available bandwidth. 
%In addition, materializing intermediate attention tensors—such as score matrices of size 128×128 per query-block–KV-block pair—would significantly increase off-chip traffic.

FAST-Prefill restructures sparse attention execution around a block-major schedule. Instead of iterating over query blocks or heads, the unit iterates over KV blocks in ascending block index order. Prior to computation, the sparse index set is transformed into a compact job list representation, where each KV block identifier $b$ is associated with a list of consumers $(h,q_b)$, indicating that attention head $h$ and query block $q_b$ require this KV block. This transformation replaces global sorting with a linear-time bucketization step. A block-use counter records the number of times each KV block is referenced and computes offsets. 
%job records are scattered into a contiguous memory region such that all consumers of a given KV block form a sequential run. As a result, the controller can traverse KV blocks sequentially, fetch each required block once as a long burst, and service all dependent attention computations while the block resides on-chip.

KV residency is managed using a liveness-driven cache replacement policy rather than a naive frequency-based policy. For each KV block $b$, the block-use counter computed during job construction serves as an exact remaining-use counter, indicating how many attention computations still require that block. Each time the block is consumed, the counter is decremented; when it reaches zero, the block is provably not required for the remainder of the sparse attention step. This property enables an evict-on-nill rule that eliminates unnecessary refetches. However, this is inefficient as the cache size is limited. To improve performance under limited URAM capacity, the cache is partitioned into two regions: a Hot KV tier and a Cold KV tier, with tags stored in BRAM. Blocks with high remaining use—i.e., those whose reuse count exceeds a static threshold $T_{hot}$
—are admitted to the hot region, while blocks with limited reuse are placed in the cold region or bypass the cache entirely. The threshold $T_{hot}$ is chosen to be 50\% of the total query blocks.
%hardware cost considerations: it represents the minimum reuse required to amortize the URAM footprint of a KV block over multiple attention computations. 
This prevents moderately reused blocks from displacing heavily reused ones and avoids cache thrashing.

Prefetching is coordinated by a lightweight local FSM that operates alongside the global controller. The local FSM maintains a bounded lookahead window over upcoming KV blocks in block-index order and consults the remaining-use counters to determine whether a block will be used and where it should be placed. Blocks with zero remaining use are skipped entirely, while blocks already resident in the cache are not refetched. Prefetch requests are issued only when sufficient space is available in the appropriate cache region, preventing premature eviction of live blocks. Because block-usage counters monotonically decrease, the prefetch and eviction decisions are tightly aligned with the execution schedule, ensuring that blocks are fetched neither too early nor too late.

Attention computation is performed using the Hybrid MPU. For each KV block fetched into the cache, the controller streams the corresponding Key tile into an on-chip buffer and iterates over its job list. For each job, the query block is read, and a score tile of size 128×128 is computed. Softmax normalization is performed in a streaming fashion, and the resulting attention weights are immediately applied to the corresponding Value tile to produce partial outputs. These partial results are accumulated directly into an on-chip output buffer indexed by $(h,q_b)$. By reducing score computation, normalization, and value aggregation into a single fused pipeline, the SAU avoids generating large tensors and confines all transient data to bounded FIFOs and buffers.

Because execution proceeds in KV-block order rather than query order, partial outputs for each query block arrive out of order. Rather than introducing a heavy reorder network, FAST-Prefill uses keyed accumulation: each job record carries the identifiers of its destination head and query block, and partial results are added into a banked accumulator memory at deterministic addresses. This mechanism functions as a reorder buffer in effect, while remaining hardware-light. Once all KV blocks have been processed and all block-usage counters reach zero, the accumulator contains the final sparse attention outputs in order and ready for FFN processing.

\subsection{Hybrid Matrix-Multiplication Unit}
\label{Sec_MPU}

Sparse attention in long-context inference requires executing a large number of block-level dot products across attention heads and query blocks. While each dot product is structurally regular, the aggregate compute demand is substantial: supporting parallel execution across heads and blocks would require instantiating many matrix multiplication engines. On FPGA platforms, using only DSP-based multipliers is inefficient due to the limited number of DSPs, severely constraining scalability and preventing parallel instantiation. This motivates a new approach to arithmetic realization that enhances throughput. To this end, we design a hybrid matrix processing unit that is based on bit-plane arithmetic in addition to DSP matrix grids, all tailored for INT-8 precision.

%—one that trades DSP usage for LUT-based computation while preserving throughput. To this end, we design a DSP-free matrix multiplication engine based on bit-plane arithmetic tailored for INT8 precision.

At a high level, an 8-bit signed integer multiplication can be decomposed into a sum of bit-level partial products. Given a value $a \in \mathbb{Z}_8$  and a value $b \in \mathbb{Z}_8$, we represent them as:
\begin{equation}
     a = \sum_{i=0}^{7}a_i.2^i, b = \sum_{j=0}^{7}b_j.2^j,
\label{eq_nib1}
\end{equation}
where $a_i,b_j \in \{0,1\}$ denote individual bit planes. The product is computed as: 
\begin{equation}
     a.b = \sum_{i=0}^{7}\sum_{j=0}^{7}(a_i \wedge b_j).2^{i+j},
\label{eq_nib2}
\end{equation}

Each partial product reduces to a simple AND operation followed by a shift, which maps naturally onto LUT logic. However, directly instantiating all $8 \times 8$ bit-plane interactions is neither area-efficient nor latency-optimal.

To reduce complexity while maintaining throughput, we further apply nibble partitioning, decomposing each 8-bit operand into high (H) and low (L) 4-bit halves (nibbles):

\begin{equation}
     a = a^{(H)}.2^4 + a^{(L)}, b = b^{(H)}.2^4 + b^{(L)},
\label{eq_nib3}
\end{equation}
The multiplication then expands as follows:
\begin{equation}
     a.b = (a^{(L)}b^{(L)}) + (a^{(H)}b^{(L)} +a^{(L)}b^{(H)}).2^4 + (a^{(H)}b^{(H)}).2^8,
\label{eq_nib4}
\end{equation}
Each term corresponds to an INT4$\times$INT4 multiplication, which can be efficiently implemented using small LUTs. This nibble-level decomposition significantly reduces latency and optimizes resources while preserving exact arithmetic semantics. %Importantly, the three resulting partial products are fixed-shifted and summed, enabling a structured and predictable accumulation path.

%Accumulation is a critical concern, as replacing DSP multipliers without addressing the accumulator would simply shift DSP pressure on accumulation. 
In our design, accumulation is implemented entirely using LUT-based adders built from FPGA carry chains. Modern FPGAs provide dedicated fast carry logic that enables multi-bit additions with single-cycle latency, comparable to DSP adders for moderate bit-widths. 
This approach enables parallel systolic arrays with each array based on either bit-plane based arithmetic or DSP based logic. On Xilinx U280 FPGA, we designed the hybrid MPU to have six 32x32 arrays based on DSPs and six 32x32 arrays based on bit-plane based arithmetic. Multiplication in the PEs compute with INT-8 precision and accumulation in INT-32. 
%Each processing element maintains a local accumulator that incrementally adds the shifted INT4 $\times$ INT4 partial products as operands stream through. Because accumulation occurs every cycle once the pipeline is filled, the bit-plane decomposition does not introduce serialization or throughput loss. The accumulator width is statically provisioned to accommodate the maximum dot-product range, ensuring correctness without overflow.

%The resulting processing element is fully pipelined with bit-plane generation, partial product computation, shifting, and accumulation being streamlined. While the arithmetic realization differs from conventional DSP-based MAC units, the logical computation remains a standard dot product. 

%This DSP-free approach enables dense instantiation of matrix multiplication engines, allowing multiple block-level computations to proceed in parallel across attention heads or query blocks. Unlike DSP-based designs, whose throughput is bounded by the fixed DSP budget, the proposed architecture scales with LUT availability, which is substantially higher on devices such as U280. This shift is particularly important for sparse attention, where performance is dictated not by a single large matrix multiply but by many concurrent block-sparse dot products driven by irregular sparsity patterns. By exploiting LUT abundance and fast carry chains, the proposed bit-plane engine sustains high throughput while freeing DSP resources entirely.

\begin{figure*}[t]
    \centering
    \includegraphics[width=0.99\textwidth]{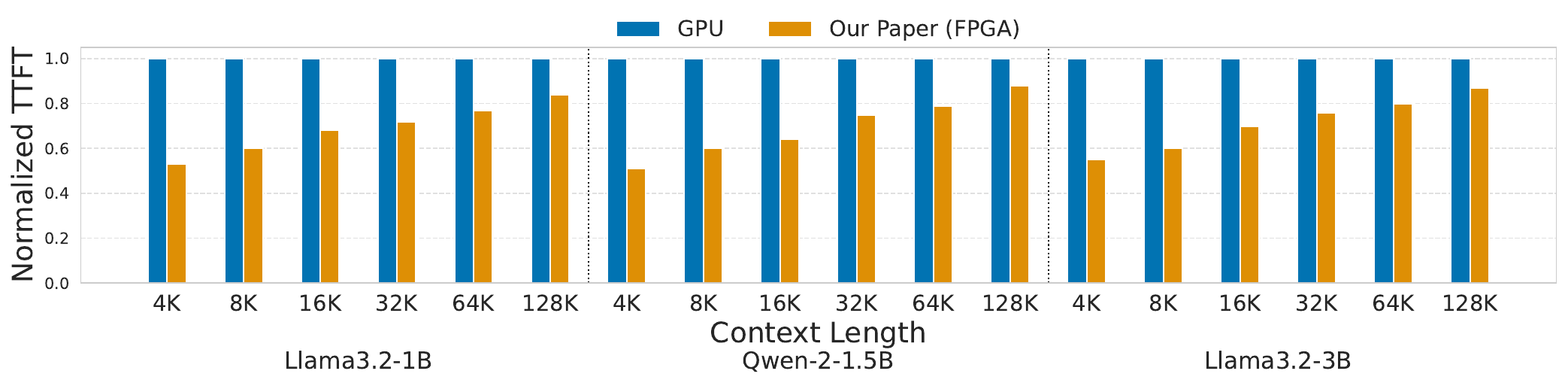}
    \caption{Comparing TTFT of FAST-Prefill with the Baseline GPU implementation}
    \vspace{-0.2cm}
     \label{fig:ttft}
\end{figure*}

\subsection{Other Hardware Units}
\label{Sec:other_units}
\subsubsection{Global Finite State Machine (FSM)} The global FSM orchestrates the execution of the entire accelerator. It tracks high-level execution phases, including chunk ingestion, KV generation, SIGU, SAU, and feed-forward processing. The FSM enforces order between stages, ensures barrier synchronization of chunks when needed, and arbitrates access to shared resources MPU. %By centralizing control, the FSM enables deterministic execution while allowing individual compute units to operate as deeply pipelined streaming units.

\subsubsection{Special Function Unit (SFU)} SFU computes operations such as softmax, normalization, and activation functions such as SiLU. This is shared across other compute units and coordinated by the Global FSM.

\begin{figure*}[t]
    \centering
    \includegraphics[width=0.99\textwidth]{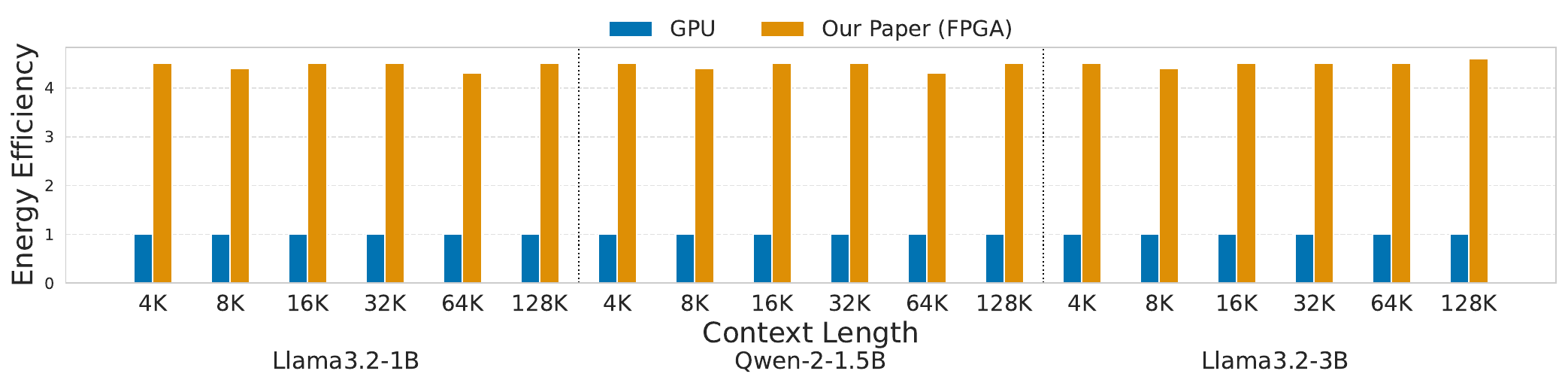}
    \caption{Comparing Energy Efficiency of FAST-Prefill with the Baseline GPU implementation}
    \vspace{-0.2cm}
     \label{fig:energy}
\end{figure*}

\begin{table}[t]
\centering
\caption{Hardware parameters of GPU and FPGA platforms.}
\label{tab:hardware_comparison}
\begin{tabular}{c|cc}
\toprule
\textbf{Platform} & \textbf{NVIDIA} & \textbf{AMD} \\
                  & \textbf{A5000 GPU} & \textbf{U280 FPGA} \\
\midrule
\textbf{Compute units} & 8,192 & 9,024 \\
                       & CUDA Cores & DSP48s \\
\textbf{Frequency (MHz)} & 1,695 & 175 (Achieved) \\
\textbf{TOPS} & 222 & 5.4 \\
\textbf{Memory (GB)} & 24 & 8 (HBM) \& 32 (DDR) \\
\textbf{BW (GB/s)} & 768 & 38 (DDR) \& 460 (HBM) \\
\bottomrule
\end{tabular}
\label{Tab:platform_Details}
\end{table}

\begin{table}[t]
\centering
\caption{FPGA Resource utilization}
\label{tab:resource_breakdown}
\begin{tabular}{lccccc}
\toprule
\textbf{Module} & \textbf{LUT (k)} & \textbf{FF (k)} & \textbf{BRAM} & \textbf{URAM} & \textbf{DSP} \\
\midrule
% \textbf{Matrix Processor} & & & & & \\
% \quad Hybrid MPU (QKV) & 525 & 852 & 128 & 0 & 8,217 \\
% %\quad LUT MPU (Wo) & 577 & 592 & 0 & 0 & 20 \\
% %\quad LUT MPU (W1) & 576 & 589 & 0 & 0 & 18 \\
% %\quad LUT MPU (W2) & 577 & 591 & 0 & 0 & 26 \\
% \midrule
% \textbf{Attention Unit} & & & & & \\
% \quad Sparse Attention & 15 & 9 & 49 & 0 & 13 \\
% \quad Index Generation & 48 & 63 & 0 & 0 & 2 \\
% \midrule
% \textbf{Activation Functions} & & & & & \\
% \quad RMSNorm (×2) & 4 & 3 & 0 & 0 & 6 \\
% \quad RoPE (×2) & 4 & 4 & 0 & 0 & 42 \\
% \quad Residual Add (×2) & 1 & 0.4 & 0 & 0 & 0 \\
% \midrule
% \textbf{Global Buffers} & 0 & 0 & 2,818 & 1,536 & 0 \\
% \midrule
\textbf{Used} & 838 & 1232 & 2250 & 912 & 6459 \\
\textbf{Available} & 1,304 & 2,607 & 4,032 & 960 & 9,024 \\
\textbf{Utilization (\%)} & 64.3 & 47.3 & 55.8 & 95 & 71.6 \\
\bottomrule
\end{tabular}
\end{table}

\begin{table}[t]
\centering
\caption{Accuracy comparison on RULER benchmark ($\uparrow$)}
\label{tab:ruler_comparison}
\resizebox{\columnwidth}{!}{%
\begin{tabular}{cccccccc}
\toprule
\textbf{Model} & \textbf{Method} & \textbf{4k} & \textbf{8k} & \textbf{16k} & \textbf{32k} & \textbf{64k} & \textbf{Avg} \\
\midrule
\multirow{3}{*}{LLaMA-3.2-1B} & FlexPrefill (BF-16) & 95.67 & 60.58 & 48.32 & 53.60 & 50.24 & 61.68 \\
 & FlexPrefill (INT-8) & 53.61 & 28.12 & 27.14 & 30.12 & 28.22 & 33.44 \\
 & \textbf{FAST-Prefill} & \textbf{52.14} & \textbf{27.33} & \textbf{27.12} & \textbf{28.50} & \textbf{30.30} & \textbf{33.07} \\
\midrule
\multirow{3}{*}{LLaMA-3.2-3B} & FlexPrefill (BF-16) & 93.51 & 79.09 & 74.76 & 73.08 & 68.27 & 77.74 \\
 & FlexPrefill (INT-8) & 73.08 & 72.60 & 62.27 & 56.55 & 52.75 & 63.45 \\
 & \textbf{FAST-Prefill} & \textbf{71.11} & \textbf{70.40} & \textbf{59.36} & \textbf{54.87} & \textbf{50.69} & \textbf{61.28} \\
\bottomrule
\end{tabular}}
\label{tab:ruler}
\end{table}

\section{Evaluation}

\begin{figure}
    \centering
    \includegraphics[width=0.99\linewidth]{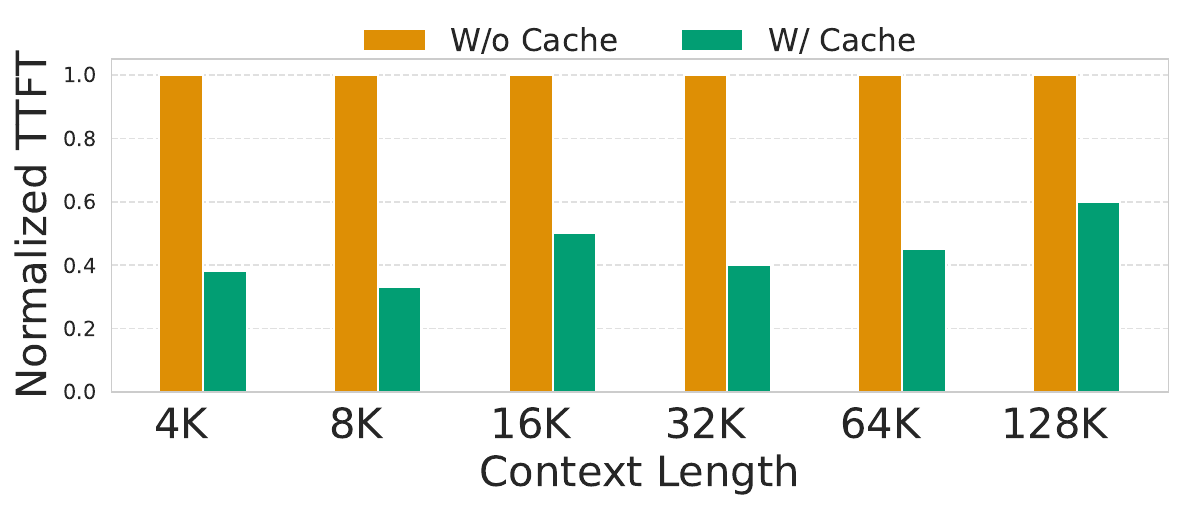}
    \caption{Ablation: Impact of Cache on TTFT [Llama-3.2-3B]}
    \vspace{-0.2cm}
     \label{fig:ttft_Cache}
\end{figure}

\begin{figure}
    \centering
    \includegraphics[width=0.99\linewidth]{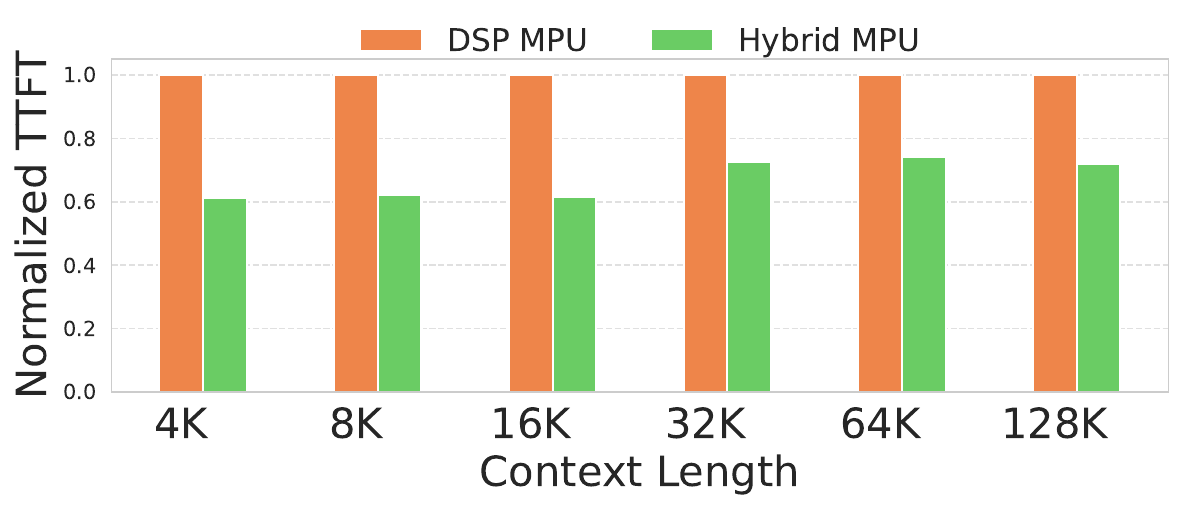}
    \caption{Ablation: Impact of Hybrid MPU on TTFT [Llama-3.2-3B]}
    \vspace{-0.2cm}
     \label{fig:ttft_mpu}
\end{figure}

\subsection{Evaluation Setup}
\textbf{Models and Benchmark.} We evaluate the effectiveness of FAST-Prefill with state-of-the-art LLMs: Llama3.2-1B-Instruct \cite{llama}, Qwen2.5-1B-Instruct \cite{qwen} and Llama3.2-3B-Instruct \cite{llama}. Following Flex-Prefill \cite{flexprefill}, we evaluate accuracy using the RULER \cite{hsieh2024ruler} benchmark. We compare FAST-Prefill against the official Flex-Prefill implementation \cite{flexprefill} baselines.

\textbf{Metrics.} We use Time To First Token (TTFT) as the latency metric for the prefill stage. We evaluate TTFT over a range of context lengths [4K, 8K, 16K, 32K, and 128K]. Additionally, we report the energy efficiency in Token/Joule [Token count = 1 in prefill]. For our evaluations, we set the batch size to 1. 

\textbf{FPGA Platform and Baselines.} We implement FAST-Prefill on Xilinx Alveo U280. Table-\ref{Tab:platform_Details} details the platform parameters. We compare the performance of FAST-Prefill against the Nvidia A5000 GPU. We evaluate FAST-Prefill (W8A8) against the quantized INT-8 Flex-Prefill implementation. 

\textbf{FPGA Implementation.} We implement our design using Vitis HLS v2023.2 and Vivado v2023.2. For latency measurements, We use the Vitis Analyzer tool for FPGA and PyTorch timing library for GPU measurements. We use the Vivado tools, nvidia-smi for energy measurements.

\subsection{Evaluation Results}

\subsubsection{Evaluation of Model Accuracy}
We detail the accuracy results from the RULER benchmark in Table-\ref{tab:ruler}. With INT-8 quantization of Flex-Prefill, the weights and activations are all quantized to INT-8. However, the matrix multiplication requires dequantization to 16 bits. With W8A8 quantization, where all computations are in INT-8, FAST-Prefill achieves accuracy similar to that of the INT-8 implementation of Flex-Prefill. 

\subsubsection{Evaluation of Performance}
We compare the latency of FAST-Prefill with the GPU implementation of Flex-Prefill. %We compare latency in two segments: Time To First Token (TTFT) for the prefill phase and Time Per Output Token (TPOT) for the decode phase.
Figure-\ref{fig:ttft} demonstrates that FAST-Prefill achieves lower TTFT than the A5000 GPU implementation of Flex-Prefill. FAST-Prefill achieves a speedup of 1.5$\times$ to 2.5$\times$ in TTFT over both the GPU implementations across varied context lengths. This speedup can be attributed to the customized kernels for sparse index generation and the liveness-driven two-tier cache, which reduce off-chip memory traffic via operation fusion and increased on-chip buffer access. GPU implementation, on the other hand, incurs frequent irregular access to its off-chip memory due to the lack of reuse-oriented, efficient prefetching of KV cache blocks. Additionally, the GPU offloads most parts of the sparse index generation logic to the CPU, which contributes to higher latency.
%During the decode stage, the attention is dense and heavily memory-bound as attention spans over all the KV cache blocks for generating a single output token. Figure-bbb demonstrates that CONFLUX achieves up to 1.5$\times$ speedup in TPOT over GPU across diverse output lengths. This speedup can be attributed to the efficient use of on-chip buffers to reduce off-chip memory traffic.

\subsubsection{Evaluation of Energy Efficiency}
We compare the energy efficiency of FAST-Prefill with GPU implementations. Figure-\ref{fig:energy} shows the energy efficiency (Token/Joule) results with FAST-Prefill achieving up to 4.5$\times$ energy efficiency over GPU implementations.

\subsection{Ablation Studies}

\subsubsection{Impact of Liveness-driven cache}
We compute the performance with and without the liveness-driven cache (16 MB) under the same hardware constraints on the remaining components. As shown in Figure-\ref{fig:ttft_Cache}, compared to the cacheless design, the design with cache demonstrates 2.5$\times$ improvement in TTFT. This difference can be attributed to the 65\% hit rate, which contributes to a significant reduction in off-chip memory access.

\subsubsection{Impact of Hybrid Matrix Multiplication Unit}
In comparison to DSP-only MPU, Hybrid MPU provides 1.8$\times$ speedup in latency as shown in Figure-\ref{fig:ttft_mpu}. Additionally, without the Hybrid MPU design, approximately 85\% of LUT resources would remain idle. 

\section{Related Work}
Recent works \cite{flightllm, speedllm, edgellm, embedfpga, lutllm, terrific, cdllm, lembda, tellme, terafly, looplynx} have explored LLM inference on FPGAs. Works such as \cite{embedfpga, terafly, cdllm} focus on enhancing decode stage performance via custom dataflow architectures. Other works such as \cite{terrific, tellme, lutllm} emphasize on extreme quantization and design kernels to exploit the quantization benefits. Previous works \cite{flightllm} focus on exploiting weight sparsity and on-chip decode to achieve higher performance than GPUs. Works \cite{looplynx} target multi-FPGA environments to scale decode compute for large models. Some works \cite{cdllm, edgellm} target CPU-FPGA heterogeneous compute. However, all the works primarily target short context lengths ($<$1024 tokens). These works do not scale well for the compute-intensive prefill stage as they all perform dense attention. Previous works on accelerating long-context inference \cite{accllm} target a single attention sparsity pattern and focus on achieving performance via weight sparsity and mixed precision. Additionally, these works target context lengths up to 8K tokens. Conversely, our work focuses on efficiently exploiting dynamic sparsity patterns in attention for very long contexts in the prefill stage. Optimizations of weight sparsity and efficient token generation in the decode stage are orthogonal to our approach and can contribute to further performance enhancements.

\section{Conclusion}
In this work, we proposed FAST-Prefill, the first FPGA accelerator to the best of our knowledge for prefill stage inference acceleration for long contexts with dynamic attention sparsity. We leverage the Flex-Prefill algorithm for sparse index generation. We design streaming sparse index generation logic with fused kernels to mitigate large tensors. FAST-Prefill features a liveness-driven two-tier cache to reduce irregular HBM accesses and promote efficient reuse of KV cache blocks. Additionally, we design a hybrid matrix multiplication unit to enhance GEMM throughput via LUT-based design. FAST-Prefill achieves up to 2$\times$ speedup in TTFT and 4$\times$ improvement in energy efficiency over the efficient GPU implementation of Flex-Prefill. Our design can be enhanced by incorporating compression techniques such as N:M and block pruning. These techniques are orthogonal to our approach and will contribute towards latency improvements. 
\bibliographystyle{IEEEtran}
\bibliography{references}


% \end{thebibliography}

\end{document}